\documentstyle[prl,aps]{revtex}
\draft

\begin{document}
\twocolumn[\hsize\textwidth\columnwidth\hsize  
            \csname @twocolumnfalse\endcsname   

\title{Two-Dimensional Fluctuations Close to the Zero-Field
      Transition of Bi$_2$Sr$_2$CaCu$_2$O$_8$}

\author{J. K\"otzler and M. Kaufmann}
\address{Institut f\"ur Angewandte Physik und Zentrum f\"ur
         Mikrostrukturforschung,\\
         Jungiusstr. 11, D -- 20355 Hamburg, Germany}

\date{July 15, 1997}
\maketitle

\widetext

\begin{abstract}
The ac-conductivity along the CuO$_2$-layers of epitaxial films
and single crystals of thicknesses 0.2$\mu$m$\leq L_z \leq$130$\mu$m
reveals clear signatures for $2D$-ordering. In particular, at $T_c$ the
screening length assumes the universal value $L_z\Lambda_{T_c}/2\nu_s$
predicted by the Kosterlitz-Thouless theory, which indicates that
binding of long vortex strings nucleates long-range superconductivity
in the present samples containing up to $10^5$ CuO$_2$-layers.
Several details of the order-parameter dynamics remain unexplained.
\end{abstract}

\hspace{0.5cm}

\pacs{PACS numbers: 74.40.+k, 74.25.Fy, 74.72.Hs}

]

\narrowtext
The layered structure of the high-T$_c$ oxide superconductors (HTSC)
promotes the formation of topological exitations which may reduce
$T_c$ to temperatures significantly
below the mean-field value $T_{c0}$, where a finite Cooper-pair density
$n_s^0$ is created. The dramatic suppression of $T_c$ to below $0.5
T_{c0}$ was observed by Rogers et al.\cite{1} in the dynamic resistance
of a single unit cell thin Bi$_2$Sr$_2$CaCu$_2$O$_8$ film and related to
the Kosterlitz-Thouless (KT) unbinding\cite{2,3} of pancake-shaped
vortex-antivortex (VA) pairs at
$T_c=T_{KT}$. On epitaxial YBaCu$_3$O$_7$\cite{4} and
Bi$_2$Sr$_2$CaCu$_2$O$_8$ films\cite{5}, a rise of $T_c$ towards
$T_{c0}$ with increasing number $N$ of layers was inferred from
current-voltage characteristics up to $N=20$ and was successfully explained
by $2D$-ordering due to the KT-pairing of VA strings piercing the films.
In the presence of a finite interlayer Josephson coupling, characterised
by the screening length $\Lambda_J =\gamma s$ ($\gamma = \lambda_c/
\lambda_{ab}$, $s=$ layer period), Josephson-strings are predicted to
bind VA-pairs of size $r>\gamma s$ (see e.g. Ref. 6). Such vortex loops
are believed to change the transition from $2D$ to $3D$ and to increase
$T_c$ to $T_c=T_{KT}(1+\pi/\ln \gamma)^2$\cite{7,?} in thick
($N\rightarrow\infty$) samples, however, this issue is still
under debate. In particular, the possibility for an interference between
VA-fluctuations and Josephson-fluxons near $T_c$ of strongly layered
materials has been considered by Horovitz\cite{9} to discuss the
transition features in the dc-resistivities $\rho_{ab}$ and $\rho_c$
of $\mu$m-thick Bi$_2$Sr$_2$CaCu$_2$O$_8$ crystals\cite{8}. More
recently, Friesen\cite{10} took into account thermal
interlayer fluctuations and predicted a divergence of $\Lambda_J$ at some
temperature $T_c^c>T_c$, which implies 3$D$ behavior below $T_c^c$.

In order to provide a deeper insight into the ordering process
in bulk ($N\gg 1$) samples of the strongly layered HTSC
Bi$_2$Sr$_2$CaCu$_2$O$_8$, where $\gamma \gtrsim 100$, we
present here a first study of the linear ac-conductivity $\sigma(\omega)$
in the highly conducting and fluctuating regime below $T_{c0}$.
By scaling analyses, we determine static screening length
$\lambda_{\pm}^2$ associated with superconducting and
normal fluctuations above and below $T_c$, respectively. For {\it
all} temperatures above $T_c$, $\lambda_+^2$ provides {\it quantitative}
evidence for $2D$-ordering in the single crystals and the epitaxial film
under study of thicknesses $L_z$ between 0.2$\mu$m and 130$\mu$m,
and the dynamical shape of $\sigma(\omega)$ is described assuming
$2D$-fluctuations of the order parameter $\Psi$. At $T_c$, the effective
penetration depth $2\lambda^2(T_c)/L_z$ agrees with the universal
length associated with the KT transition. The high, universal value of
the phase angle at $T_c$, $\sigma''/\sigma'=13.5(1)$, is consistent with
noise data on Bi$_2$Sr$_2$CaCu$_2$O$_8$\cite{27} but is much larger than
$\sigma''/\sigma'=3.4$\cite{k1,nak} obtained on the less anisotropic
YBaCu$_3$O$_7$ ($\gamma \approx 6$). In the present samples containing
up to $10^5$ CuO$_2$-planes, signatures of $3D$-ordering appear only at
temperatures below $T_c$.

Our inductive method for determining high ac conductivities from the
measured linear magnetic susceptibility $\chi(\omega)=\chi'+i\chi''$
has been outlined before for bulk\cite{11} and thin film\cite{k1} samples.
Exact numerical relations between $\chi(\omega)$ and
$\sigma(\omega)=\sigma'+i\sigma''$ were derived by
E.H. Brandt for disks of arbitrary
thickness\cite{13}, and using a complex inversion scheme,
we evaluated $\sigma(\omega)$. We examine
here a 0.2$\mu$m thick film epitaxially grown for SQUID
fabrication\cite{15} and crystalline discs of thicknesses between 13$\mu$m
and 130$\mu$m, which were cleaved from the same crystal grown by L.
Winkeler\cite{16}. One of the great advantages of investigations of
$\sigma(\omega)$ lies in the possibility of a precise determination of
the onset for long-range superconductivity at the
temperature where the frequency variation of the phase angle,
$d(\sigma''/\sigma')/d\omega$, changes its sign\cite{17}.

The first of such results for Bi$_2$Sr$_2$CaCu$_2$O$_8$ are depicted in
Fig.1 showing well defined intersections of the phase, i.e.
$d(\sigma''/\sigma')/d\omega=0$. Several novel features are seen.
Despite the rather different $T_c$'s of crystal and film, both attain
there the same value of the phase angle, $\varphi_c\equiv
\arctan (\sigma''/\sigma')_{T_c}=85.9^0(8)$, in zero
magnetic field. Since we observed the same value also on many other
Bi$_2$Sr$_2$CaCu$_2$O$_8$ crystals of different origin\cite{22},
we consider $\varphi_c$ to
be characteristic for this strongly layered superconductor.
For the epitaxial film Fig. 1 demonstrates that an external field as
low as 0.1mT destroys the crossing of the phase angles, and that the
positive,
'paraconductive' slope, $d\varphi/d\omega>0$, remains  at all lower
temperatures. This implies the destruction of true superconductivity
by very small external fields, which is in sharp contrast to
YBa$_2$Cu$_3$O$_7$, where $3D$-ordering characterized by
$\varphi_c=75^0(2^0)$, between zero field\cite{nak} and 19T\cite{20} has
been observed using the crossing point criterion.
We regard this sensitivity of $T_c$ against externally induced
vortices as a first indication of a lowered dimensionality. For layered
materials with $\Lambda_J=0$, Martynovich\cite{Martynovich} predicted that
fluctuations of pancake vortices induced by a field as low as 1mT,
destroy $2D$-superconductivity.

In order to analyse the transition behavior in more detail, we search
for the dynamic scaling of the phase and the modulus of $\sigma(\omega)$
expected near the continuous transition\cite{17}. For the
paraconductive side, Fig. 2 in fact demonstrates that after scaling of the
frequency by a relaxation rate $\tau^{-1}$, and then by scaling of the
modulus to the dc-conductivity $\sigma_+(0)$, two well defined master
curves are
obtained which differ slightly in shape
for the film and crystal.

First we discuss the scaling parameters
displayed in Fig. 3. As the characteristic feature between $T_c$ and
$T_{c0}$\cite{meissner}, the static conductivity, displays a steep
Arrhenius-type increase extending over more than 12 decades.
One of the most interesting results from the scaling is the fact
that the relaxation times obey essentially the same strong temperature
variation as the dc-conductivities. As shown in the lower panel of Fig. 3,
this implies for the 'paraconductive' penetration depth,
$\lambda_+^2\equiv \tau/\mu_0\sigma_+(0)$, which basically measures the
length across which a superconducting fluctuation diffuses during its
lifetime $\tau$, to be independent on temperature. This feature
constitutes a qualitative signature of $2D$-ordering, in contrast to the
divergence of $\lambda_+^2\sim\xi(T)$ observed in YBa$_2$Cu$_3$O$_7$ at
$B=0$\cite{nak}, which characterizes $3D$-$XY$-type ordering\cite{17}.

The other important feature is the linear increase of
$\lambda_+^2$ with the sample thickness, $\lambda_+^2=\Lambda_+L_z$
(see Fig. 4). Here
we compare the experimental
length $\Lambda_+=10.7(5)$mm to the theoretical results taking
$2D$-Gaussian fluctuations of $\Psi$ into account\cite{19},
$\Lambda_+^G=4\Lambda_T$, or considering the topological VA
fluctuations, $\Lambda_+^V=C_1 4\pi^2\Lambda_T/7$\cite{3}. Insertion
of the thermal length $\Lambda_T=\phi_0^2/4\pi\mu_0T=2$cm$/T(K)$, yields
$C_1=8.0(1)$, roughly consistent with our results,
whereas $\Lambda_+^G$ underestimates $\Lambda_+$ by a factor of 14.
We do not see any effect of the proposed onset of a Josephson
coupling at $T_c^c>T_c$\cite{10}. 

Very striking evidence of $2D$-ordering stems from the data for
$T\leq T_c$, which for the present low frequencies can be well described by
the empirical form
\begin{equation}
\sigma_-(\omega,T)=\sigma(\omega,T_c)+i/(\mu_0\omega\lambda_-^2),
\end{equation}
with $\sigma(\omega,T_c)=\sigma_0 (i\omega\tau_0)^{-x}$, where $x=
\varphi_c/\frac{\pi}{2}$. For the dominant, imaginary part of $\sigma_-$
the validity of Eq. (1) is illustrated in Fig. 4a, where the results for
the inverse screening length are shown. As exemplified for three samples,
there exist well defined $\omega\rightarrow 0$ finite values of
$\omega(\sigma_-''(\omega,T_c)- \sigma''(\omega,T_c))$ which determine the
phase-locked, superfluid density $n_s(T)=m_e/\mu_0e^2\lambda_-^2$.
The full lines explain the
increase of $n_s(T)$ towards lower temperatures obeying the relation
$\lambda_-^{-2}(T) =
\lambda_-^{-2}(T_c)\left(1+\beta(1-T/T_c)^{\nu}\right)$.

The most interesting piece of information is provided by
$\lambda_-^{-2}(T_c)$ also plotted in Fig. 4b. Encouraged by the
number of signatures for $2D$-ordering at $T>T_c$ realized above, we
compare these data with the predictions of the KT theory\cite{3} for the
screening length at the unbinding transition, $\lambda_-^2(T_c)=
L_z\Lambda_{T_c}/2\nu_s$, assuming the maximum and a reduced value of the
stiffness of $\Psi$, $\nu_s=n_s/n_s^0$\cite{6}. Obviously, the
line with $\nu_s=1$ hits the value of the film exactly, whereas
$\nu_s=0.15$ explains the thickness dependence of
$\lambda_-^{-2}(T_c,L_z)$ for the crystal {\it
quantitatively}. Investigations on other crystals\cite{22}
indicate some correlation of $\nu_s$ with the transition width, i. e.
with the disorder in the crystal, while all the 2$D$-features
reported here are essentially unchanged.

This robust nature of the KT transition against disorder seems to be a
general property, at least we are not aware of any observation of a KT
transition in a clean superconductor. This evidence for
coherent KT ordering of long vortex strings piercing up to $N=10^5$
CuO$_2$-planes is also consistent with the estimate for the maximum
sample thickness, $L_z^{max}= \gamma^2 s \approx 10^5 s$, which still
favors VA-string pairs over rings. However, we should not overlook two
discrepancies. The first one
refers to the Arrhenius-like temperature variation of the
dc-conductivity, which turns out to be much steeper than that predicted
by the KT-theory,
$\sigma_+^V(T)\approx\sigma(T_{c0})\exp[2(b(T_{c0}-T)/
(T-T_c))^{1/2}]$\cite{3}, with $b=1.5$. However, the existing ac-data on
prototype HTSC KT-systems, like the $N=1$ layer
Bi$_2$Sr$_2$CaCu$_2$O$_8$\cite{1}
and YBa$_2$Cu$_3$O$_7$\cite{Gasparov}, also reveal thermally activated
behavior of the characteristic time, $\tau\sim
\exp(U_0/T)$, which has been related to a thermal hopping of vortices
and antivortices originating from the pristine disorder in the layered
materials. In the present samples, we are confronted with a rather
large value of the barrier $U_0=8(1)$eV, as compared to 0.1eV in the
single layer systems\cite{1,Gasparov}. Surprisingly $U_0$ {\it
does not depend} on $L_z$ or on stiffness, which is one ot the great
challenges
for a further understanding of this novel KT-type ordering in bulk
Bi$_2$Sr$_2$CaCu$_2$O$_8$.

The other discrepancy concerns to the exponents $\nu=1.0(1)$ and 1.3(1)
which describe the evolution of the superfluid density
 $n_s(T<T_c)$ in the crystals and the
film, respectively, as compared to the KT-value $\nu_0=0.5$\cite{3}.
Based on similarly large values of $\nu$ measured near the vortex-glass
transitions in YBa$_2$Cu$_3$O$_7$\cite{k1,nak}, we
attribute these exponents to $3D$-ordering in the presence of structural
disorder. Towards lower temperatures, $\lambda_-^2(T)$ of the film
tends to the bulk penetration depth $\lambda_L(0)=260$nm\cite{26},
whereas in the crystal this crossover to
bulk-screening commences at some lower temperature, $T=0.94T_c$ (
inset to Fig. 4a). We argue here that with decreasing temperature the
distance $r$ between the VA-string pairs shrinks and that at some
temperature $T^*<T_c$ they decompose into small $2D$
VA-pairs\cite{4,5,7} with a coupling energy
$U_p=k_BT\Lambda_Td/\lambda_-^2(T)\ln r/\xi$. Their dissociation
may be the reason for the KT-type features that have been
observed in early investigations of I-V characteristics on thin
Bi$_2$Sr$_2$CaCu$_2$O$_8$ crystals\cite{Artemenko}.

Let us finally emphasize two {\it dynamic\/} features of the $2D$-ordering in the
strongly layered Bi$_2$Sr$_2$CaCu$_2$O$_8$ emerging from our data.
The exponent of the characteristic power law
{\it at $T_c$}, $\sigma(\omega,T_c)\sim(i\omega\tau_0)^{-x}$ with
$x = 0.955(5)$, differs significantly from that observed
for the $3D$-ordering at $B=0$ in YBa$_2$Cu$_3$O$_7$,
$x = 0.84$, \cite{nak} and also
from the mean-field result $x=(4-D)/2$\cite{17}. Looking at existing
noise spectra near $T_c$ of Bi$_2$Sr$_2$CaCu$_2$O$_8$ crystals, we find
from Ref. 11 $S_{\phi}(\omega)\sim \omega^{-0.95}$, which due to the
fluctuation dissipation theorem for the present $2D$ case, $S_{\phi}\sim
\sigma'(\omega)$\cite{29}, is fully consistent with our results. Note
that $x=0.955(5)$ implies a small but distinct deviation from the
high frequency $1/\omega$ noise predicted for a homogeneous $D=2$
superconductor, which suggests to attribute this difference to the
layered structure.

In order to describe also the dynamic shape of $\sigma(\omega)$ above
$T_c$, we tried several models, including the recent extension\cite{Cap}
of the socalled Minnhagen phenomenology\cite{29}, by which we are not
able to describe the extremely sensitive phase $\sigma''/\sigma'$ (Fig. 2).
We achieved an almost ideal fit of the scaling function (see solid
curves in Fig. 2) using the ansatz:
\begin{equation}
\sigma_+(0)/\sigma_+(\omega\tau)=S_V^{-1}(\omega\tau)+\alpha
S_G^{-1}(\omega\tau)+S_c^{-1}(\omega\tau)
\end{equation}
with $\alpha=0.22$ and 0.83 for the film and the crystal, respectively.
The first term considers the dynamics of the topological VA fluctuations
for which we take from Ref. 3,  $S_V^{-1}(\tilde{\omega})=1+i\tilde{\omega}
\int_{\sqrt{\tilde{\omega}_0}}^{\sqrt{\tilde{\omega}}} \epsilon '
dy/\left((1+iy^2)\sqrt{\tilde{\omega}_0}\right)$, assuming a constant
density of state of the relaxing bound VA-pairs, $\epsilon'=1$. The
second term accounts for $2D$-Gaussian fluctuations\cite{19},
$S_G(\tilde{\omega})=\sigma_G(0)[\pi-2\arctan \tilde{\omega}^{-1} -
\tilde{\omega}^{-1} \ln(1+\tilde{\omega}^2)]/\tilde{\omega}$. To date we
have no convincing argument for the weights $\alpha$, but note that this
is the only parameter of the fits to both shape functions, and that
the shape of the phase is extremely sensitive to an even slight variation
of the model. Rather generally speaking, this ansatz implies that
gaussian fluctuations dominate on smaller length scales $r_{\omega}=
1/\sqrt{\sigma_+(0)\omega}$, whereas the topological VA fluctuations
prevail on the  large scales. This observation
is also consistent with our main conclusion that the VA binding is driving
the transition in Bi$_2$Sr$_2$CaCu$_2$O$_8$. The third term can be well
approximated by the citical power law, $S_c(\tilde{\omega}) = s_c \cdot
(i\tilde{\omega})^{-x}$, which was introduced in Eq. (1), and the origin
of which also remains to be explained.

In summary, the screening length $\lambda(T\geq T_c)$ provides
quantitative evidence for a KT-transition at $T_c$ of bulk
Bi$_2$Sr$_2$CaCu$_2$O$_8$ samples comprising up to $10^5$ CuO$_2$
planes. Below $T_c$, signatures of $3D$-ordering are observed. The
interpretation in terms of a macroscopic $2D$-ordering
phenomenon is supported by the dynamical shape of $\sigma(\omega)$ above
$T_c$, containing contributions of both topological and $2D$-Gaussian
fluctuations of the order parameter, and by the destruction of the
thermodynamic transition in extremly low magnetic fields.

We are very much indebted to T. Amrein (J\"ulich) for providing us
with the film and to L. Winkeler (Aachen) for
the crystals. We thank E.H. Brandt (Stuttgart), G. Nakielski and
H. Schmidt (Hamburg), and
K.H. Fischer (J\"ulich) for valuable discussions
and the {\it Graduiertenkolleg 'Physik nanostrukturierter Festk\"orper'}
for financal support.

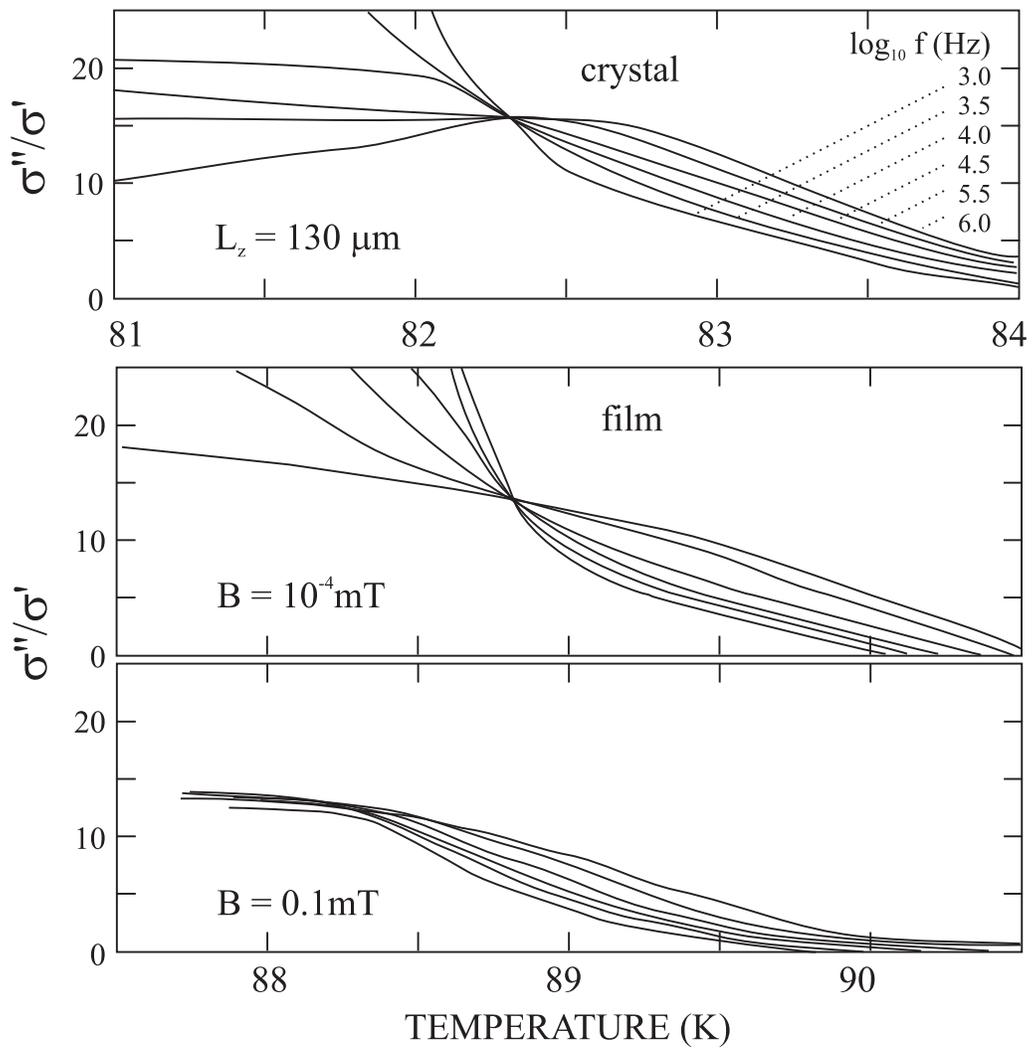
\begin{figure}
\caption{Temperature variation of the phase of the linear ab-plane
ac-conductivity of a single crystal and an epitaxial
film at vanishing magnetic field. The lower panel demonstrates the
effect of a small field. The intersection points define the
critical temperatures $T_c$.}
\end{figure}

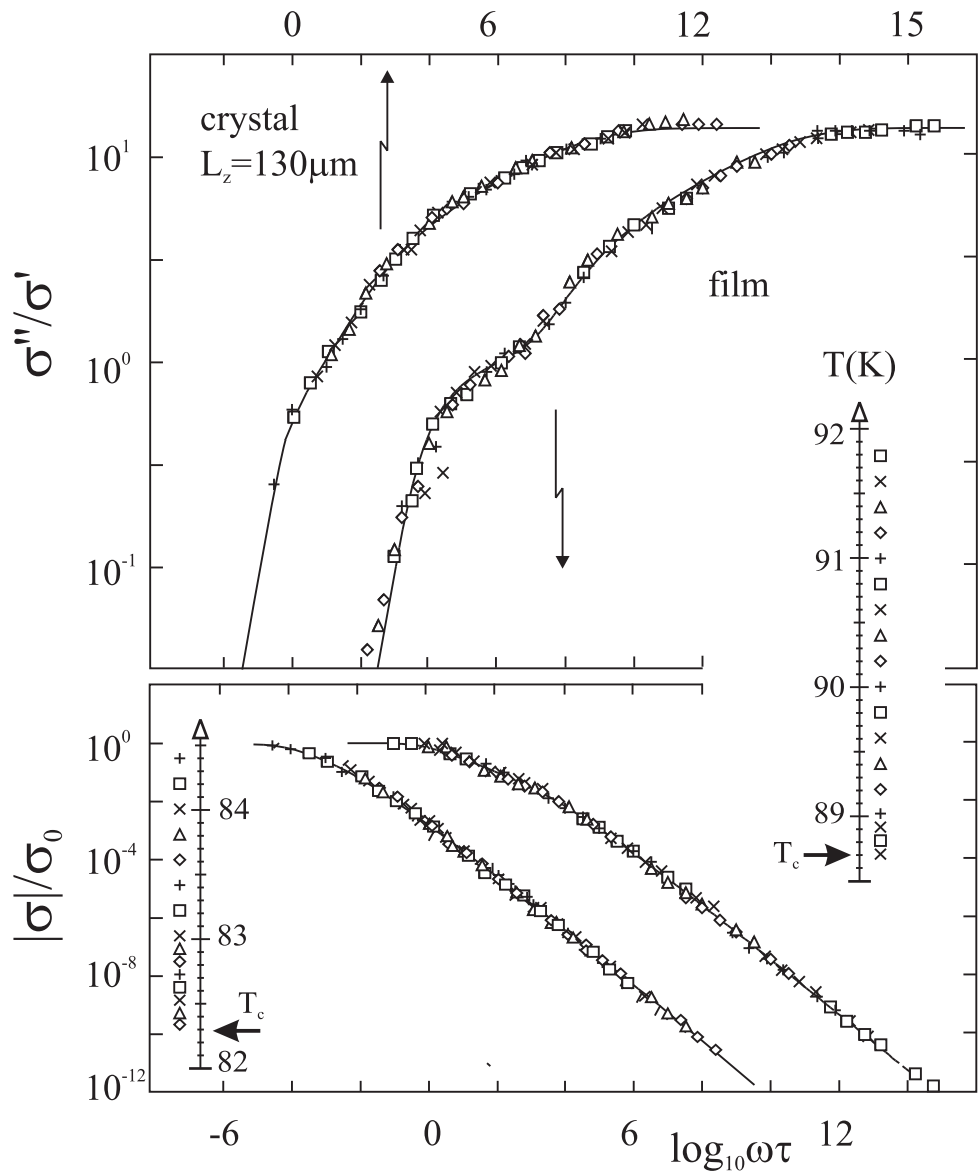
\begin{figure}
\caption{Dynamical scaling a) of the phase  and b) of the modulus of
$\sigma_+(\omega)$ above $T_c$.
Solid curves through the data were calculated from Eq. (2).}
\end{figure}

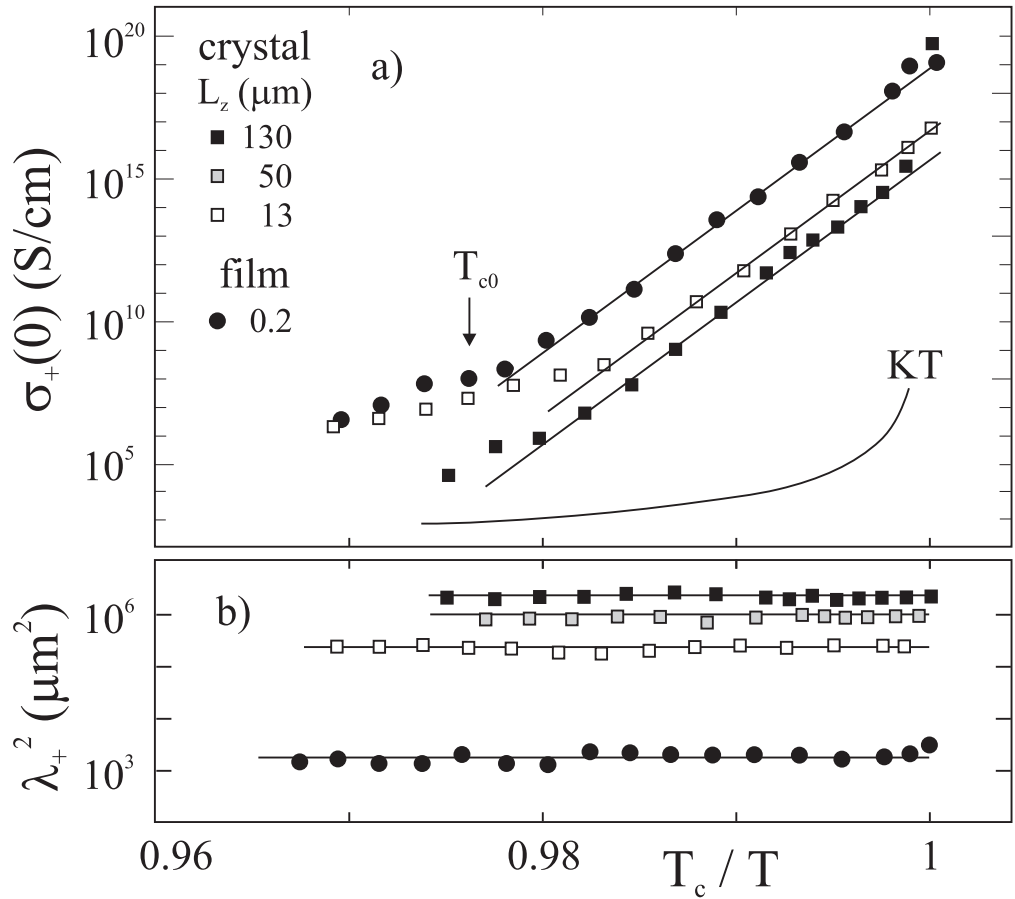
\begin{figure}
\caption{a) Arrhenius plots of the dc-conductivity $\sigma_+(0)$;
the full lines compare the experimental Arrhenius behavior to
the results of the KT-theory\protect\cite{3}.
b) Screening length
$\lambda_+^2=\tau/\mu_0\sigma_+(0)$ above $T_c$.}
\end{figure}

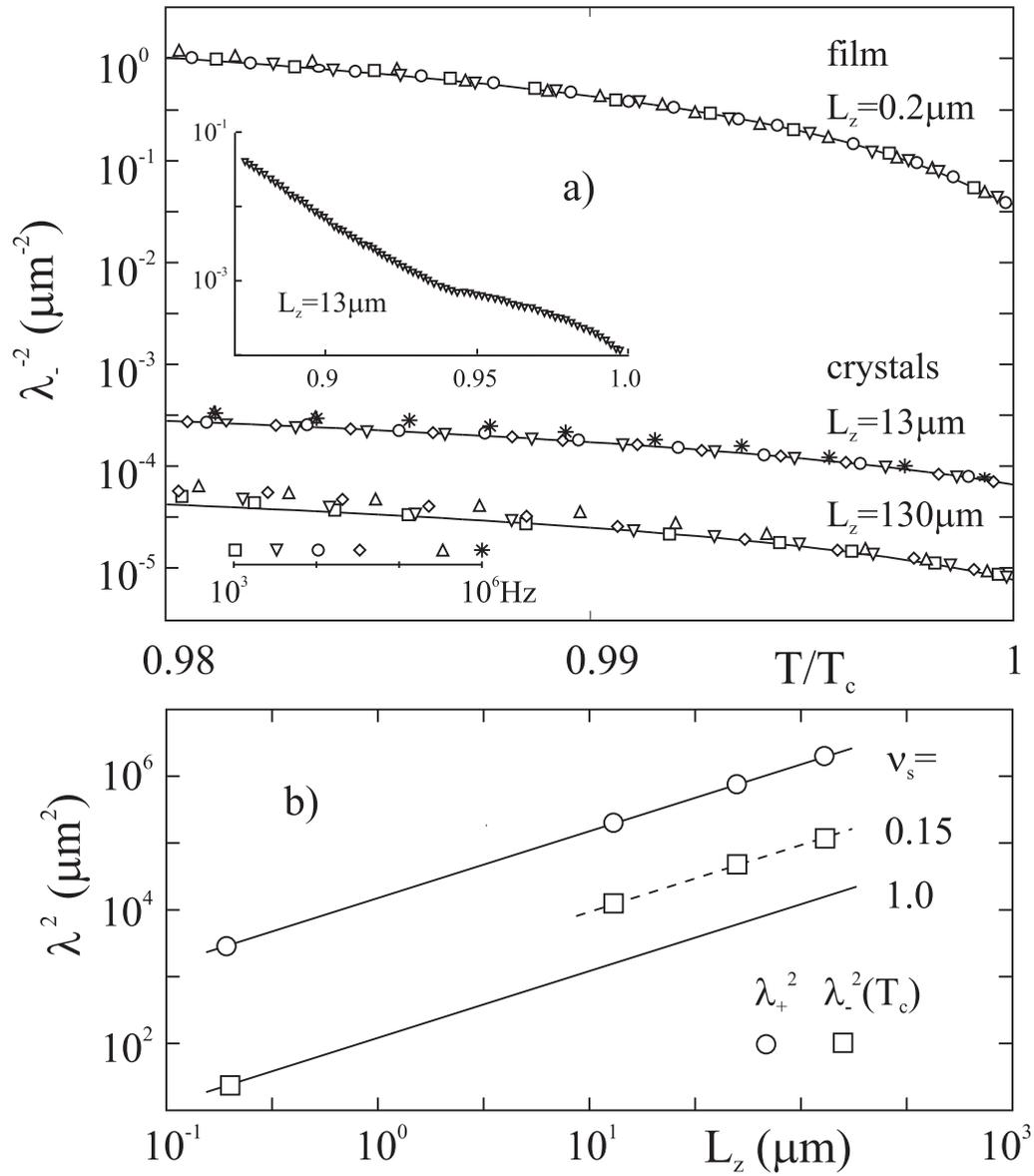
\begin{figure}
\caption{a) Squared penetration depths,
determined from the $\omega\rightarrow 0$ limit of the dc-screening
response below $T_c$, $\lambda_-^{-2}(T) \equiv \mu_0\omega (\sigma''
(\omega,T)-\sigma''(\omega,T_c))$. Solid lines are fits to Eq. (1).
b) Effect of the sample thickness $L_z$ on $\lambda_{\pm}^2$ compared
to the predictions for the KT-transition,
$\lambda_{\pm}^2=C_{\pm}\Lambda_T L_z$ with $C_{\pm}$
discussed in the text.}
\end{figure}

\end{document}